\documentclass[twoside]{dis07}
\usepackage[latin1]{inputenc}
\usepackage[dvips]{graphicx,epsfig,color}
\usepackage{wrapfig,rotating}
\usepackage{amssymb,amsmath,array}

\begin{document}
\title{Critical tests of unintegrated gluon distributions }

\author{H. Jung$^1$, A.V. Kotikov$^2$,
 A.V. Lipatov$^3$ and N.P. Zotov$^3$
%
\vspace{.3cm}\\
%
1- DESY, Hamburg, Germany\\
%
\vspace{.1cm}\\
2- BLTHPH - JINR, Dubna, Russia\\
\vspace{.1cm}\\
3- SINP - MSU, Moscow, Russia \\
}

\maketitle

\begin{abstract}
We use the unintegrated Parton Density Functions of the gluon
obtained from a fit to measurements of the structure functions $F_2(x, Q^2)$ and $F_2^c(x,Q^2)$ at HERA  to
describe the experimental data for $F_2^b(x,Q^2), F_L(x,Q^2)$
and $F_L$ at fixed $W$.
\end{abstract}

\section{Introduction}
The purpose of the present investigation is to study the longitudinal structure function (SF) $F_L (x, Q^2)$ as well as the charm and beauty contributions 
to the proton SF $F_2(x, Q^2)$ using the $k_T-$factorization approach of QCD
\cite{CCH}. The SF $F_L (x, Q^2)$ is directly connected to the gluon density in the proton. Only in
the naive quark-parton-model $F_L (x, Q^2)=0$, and becomes non-zero in pQCD.
However the pQCD leads to controversal
results still. It was shown recently~\cite{Thorne}, that the $F_L$ experimental data from HERA seem to be inconsistent with some of the 
NLO predictions (in particular the MRST one) at small $x$. BFKL
effects significantly improve the description of the low 
$x$ data when compared to a standard NLO $\bar{MS}$-scheme global 
fit. The NNLO global fit becomes better when taking into account 
higher order terms involving powers of $\ln (1/x)$. It means, that we need a resummation procedure. 

On the other hand it is  known, that the BFKL effects are taken into account from the very beginning in the $k_T-$factorization approach~\cite{CCH},
which is based on the BFKL~\cite{BFKL} or CCFM~\cite{CCFM} evolution equations summing up the large logarithmic terms
proportional to $\ln (1/x)$ or  $\ln (1/(1-x))$  in the LLA.
Some applications of the $k_T-$factorization approach were shown in Refs.~\cite{Smallx}.
In the framework of $k_T$-factorization the
study of the longitudinal SF $F_L$ began already ten years ago~\cite{CH},  where the small $x$ asymptotics of $F_L$ has been evaluated, using the BFKL results. Since 
we want to analyze the SF data in a broader range at small $x$
we use a more phenomenological approach in our analyses of $F_2$
and $F_L$ data~\cite{BKS,KLZ}. 
Using the $k_T$-factorization approach for the description of different SF 
at small $x$ we hope to obtain additional information (or restrictions), in particular, about one of the main 
ingradient of $k_T$-factorization approach - the unintegrated 
gluon distribution (UGD) 

In the $k_T$-factorization
the SF $F_{2,L}(x,Q^2)$ are driven at small $x$ primarily
by gluons and are related in the following way to the UGD 
$x{\cal A}(x,{\mathbf k}_{T}^2,\mu^2)$
\begin{eqnarray}
F_{2,L}(x,Q^2) ~=~\int^1_{x} \frac{dz}{z} \int^{Q^2}   
dk^2_T \sum_{i=u,d,s,c} e^2_i \hat C^g_{2,L}(x/z,Q^2,m_i^2,k^2_T)x{\cal A}(x,{\mathbf k}_{T}^2,\mu^2).
\end{eqnarray}
The functions $\hat C^g_{2,L}(x,Q^2,m_i^2,k^2_T)$
can be regarded as  SF of the
off-shell gluons with virtuality $k^2_T$ (hereafter we call 
them {\it hard structure functions }). 
They are described by the sum 
of the quark box (and crossed box) diagram contribution to the
photon-gluon interaction. 

To apply Eq.(1) for SF at low $Q^2$ we change the low $Q^2$ asymptotics of
the QCD coupling constant within hard structure functions.
We have used the so called "freezing" procedure in the "soft" form,
when the argument of the strong coupling constant
 is shifted from $Q^2$ to $Q^2 + M^2$~\cite{NZ}. Then $\alpha_s = 
\alpha_s(Q^2 + M^2)$. For massless quarks  $M=m_{\rho}$ and for massive ones with  mass $m_Q, M=2m_Q$.

To calculate the  SF $F_2^{c,b}$ and $F_{L}(x,Q^2)$ we used the hard SF $\hat C^g_{2,L}(x,Q^2,m^2,k^2_T)$ from Ref.~\cite{KLZ,KLZ2}\footnote{There is full
agreement of our results with the formulae for the photoproduction of heavy quarks
from Ref.~\cite{CCH2}.} and two UGD ${\cal A}(x,{\mathbf k}_{T}^2,\mu^2)$ obtained in our previous paper~\cite{JKLZ}. These UGD are determined  by a convolution
of the non-perturbative starting distribution ${\cal{A}}_0(x)$
and CCFM evolution denoted by
$\bar{\cal A}(x,{\mathbf k}_{T}^2,\mu^2)$:
\begin{eqnarray}
x{\cal A}(x,{\mathbf k}_{T}^2,\mu^2)~=~\int dz {\cal A}_0(z)
 {x\over z} {\bar{\cal A}}({x\over z},{\mathbf k}_{T}^2,\mu^2),
\end{eqnarray}
\begin{wrapfigure}{r}{0.5\columnwidth}
\centerline{\includegraphics[width=0.45\columnwidth]{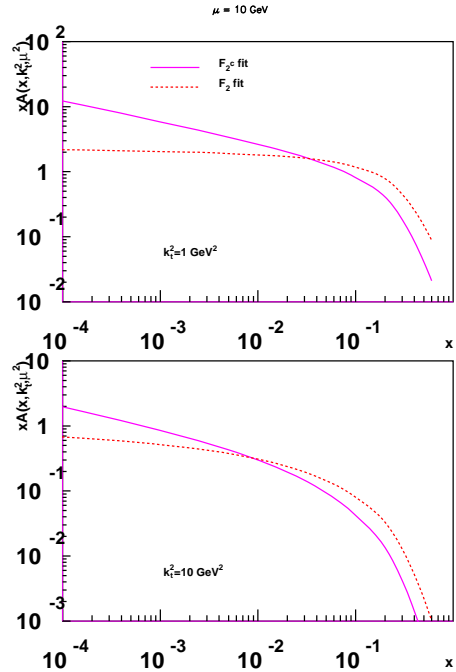}}
\caption{\it UGD obtained in the fits to $F_2^c$ (solid curve) and 
$F_2$ (dotted curve) }\label{Fig:1}
\end{wrapfigure}
where
\begin{equation}
x{\cal{A}}_0(x)~=~Nx^{-B_g}(1-x)^{C_g}(1-D_{g}x).  
\end{equation} 
 The parameters $N, B_g, C_g, D_g$ of ${\cal A}_0$ were
determined in the fits to $F_2$ and $F_2^c$ data~\cite{H1,H12} independently (see~\cite{JKLZ}) Fig. 1
shows the two different UGD.
The small $x$ behaviour of these UGD is very different\footnote{See also
Ref.~\cite{HJ}.}.

To calculate the SF $F_2^b(x, Q^2)$ and $F_L(x,Q^2)$
we took $m_c=1.4$ GeV and $m_b=4.75$ GeV
and used  the $m^2 = 0$ limit of the above Eq. 1 to evaluate the corresponding lightquark contributions to the $F_L$.
Fig. 2 (left panel) shows the $F_2^b$ as a function of $x$ at 
fixed $Q^2$.
 Fig.2 (right panel) shows the $F_L$ as a function of $x$ at fixed 
$Q^2$.
Fig. 3 shows the SF $F_L(Q^2)$ at fixed $W$ compared to the H1
data~\cite{H14}.
It is interesting to observe, that the measured $F_2^b$ seems to prefer the UGD obtained from the fit
to $F_2$ and is inconsitent with the one obtained from $F_2^c$. Also the measured $F_L$ is better
described with the UGD from the $F_2$ fit.
\begin{figure}
\hspace*{0mm}
\begin{minipage}[h]{.49\textwidth}
\includegraphics[height=8.0cm,width=7.0cm,angle=0]{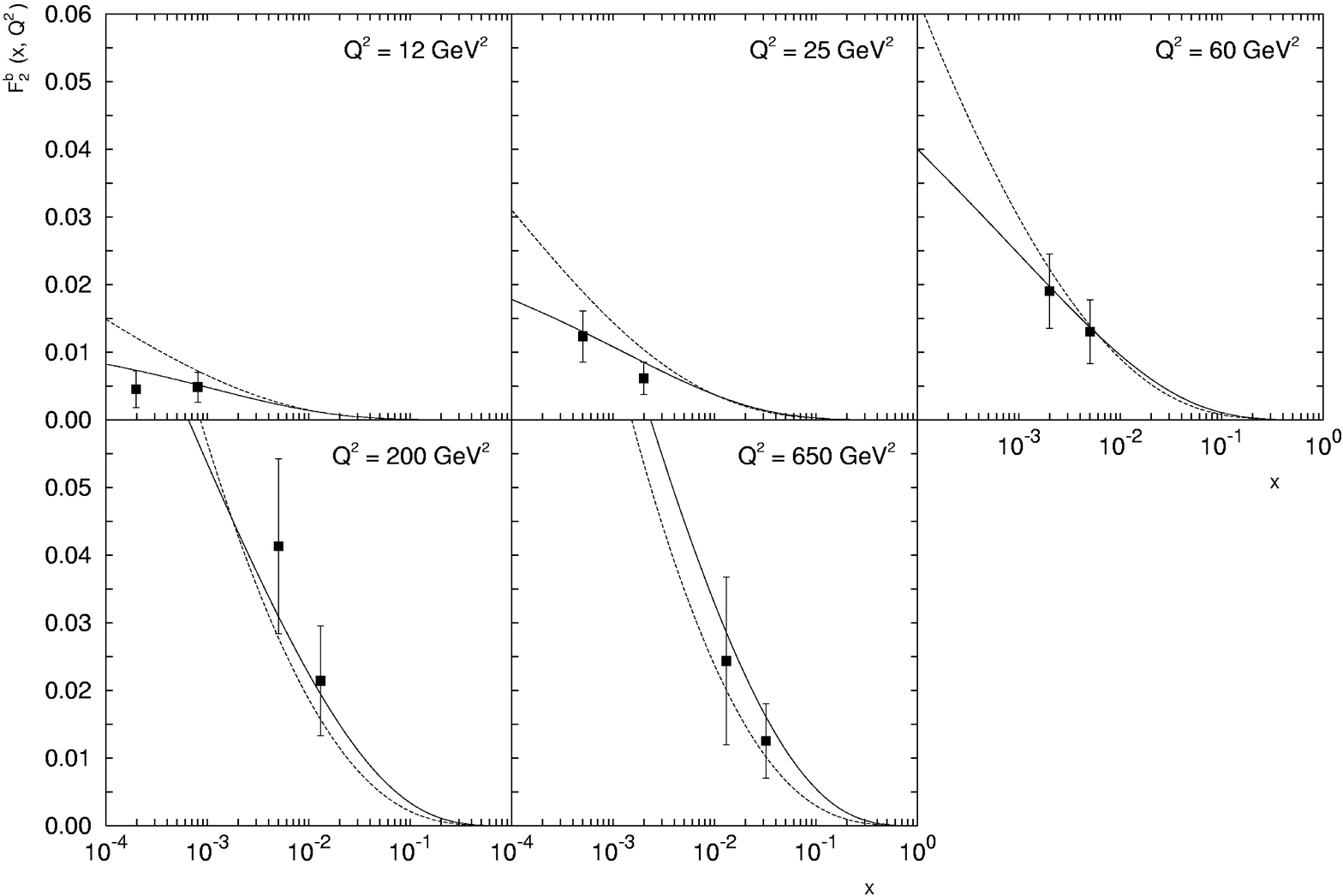}
\end{minipage}\hspace*{2mm}
\begin{minipage}[h]{.49\textwidth}
\includegraphics[height=8.0cm,width=7.0cm,angle=0]{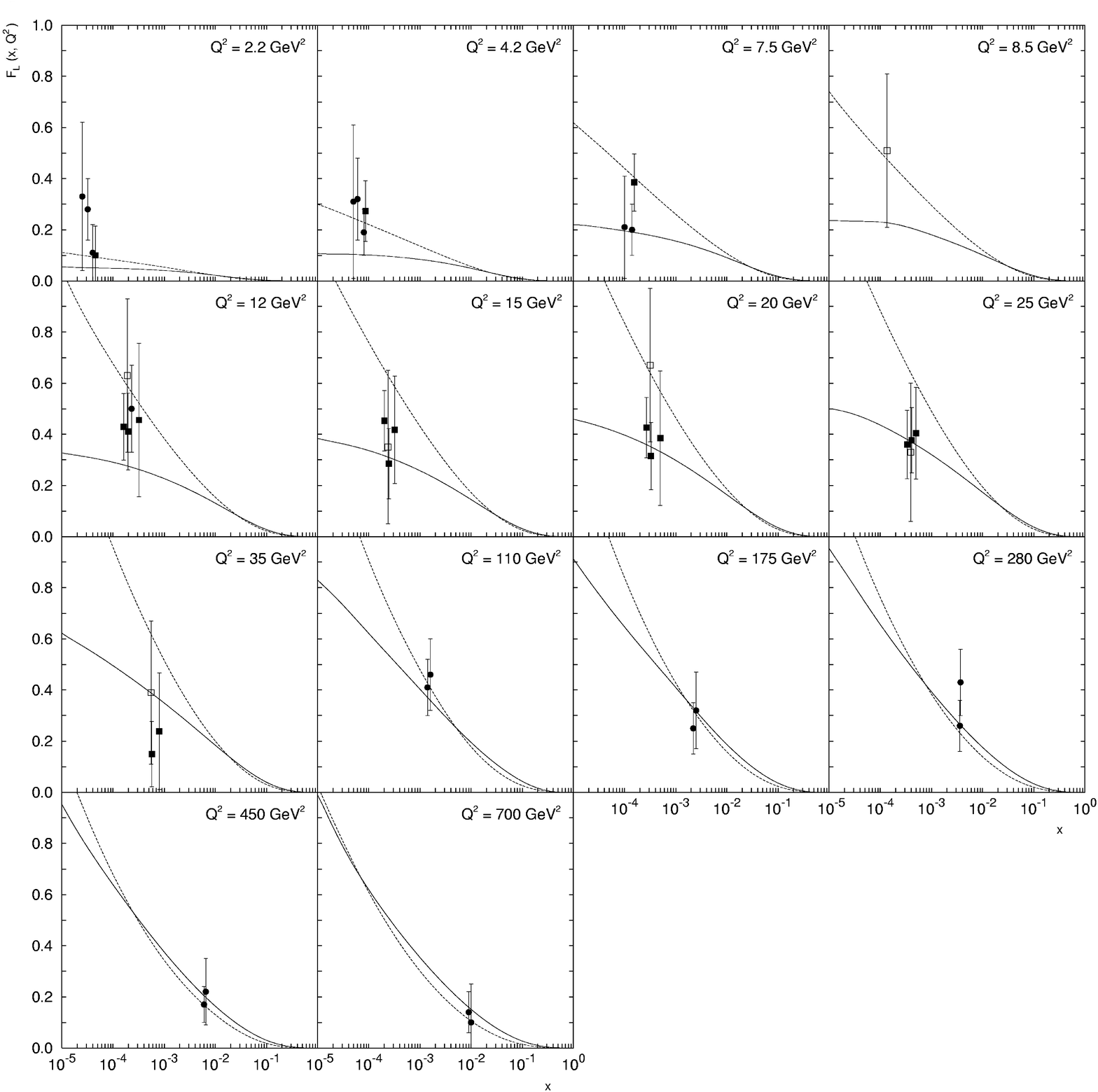}
\end{minipage}
\vspace*{-5mm}
\vspace*{-5mm}
\begin{center} 
\caption{\it The SF $F_2^b$ as a function of $x$ at
fixed $Q^2$ compared to the H1 data~\cite{H12}(left panel) 
 The solid and dotted lines
are from CCFM-evolved UGD obtained from the fits to 
$F_2(x,Q^2)$ and $F_2^c(x,Q^2)$. The SF $F_L$ as a function of $x$ 
at
fixed $Q^2$ compared to the H1 data~\cite{H1,H13}(right panel) 
 The solid and dotted lines
are from CCFM-evolved UGD obtained from the fits to 
$F_2(x,Q^2)$ and $F_2^c(x,Q^2)$.}\label{Fig:2}
\end{center}
\end{figure}
\begin{figure}
\includegraphics[height=5cm,width=13cm]{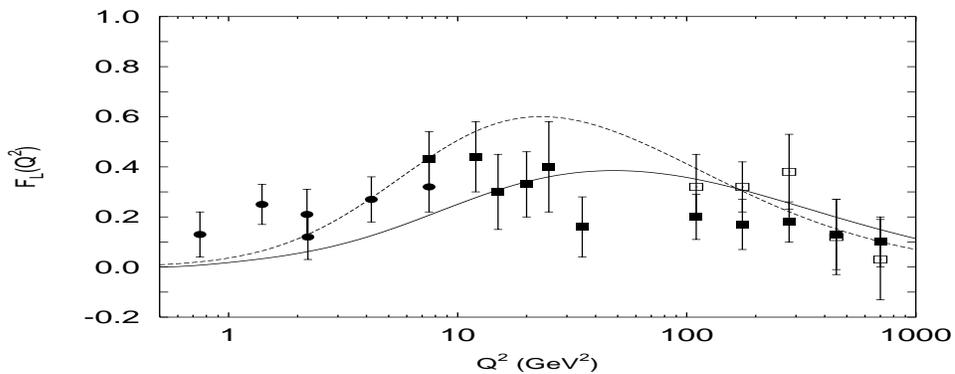}
\begin{center}
\caption{\it The $Q^2$ dependence of SF $F_L(Q^2)$ at
fixed $W = 276$ GeV compared to the H1 data~\cite{H14}
 The solid and dotted lines
are from CCFM-evolved UGD obtained from the fits to
$F_2(x,Q^2)$ and $F_2^c(x,Q^2)$.}\label{Fig:3}
\end{center}
\end{figure}
In summary the $k_T-$ factorization approach with the CCFM-evolved UGD
obtained from the fits to the $F_2(x, Q^2)$ 
data reproduces the H1 data for SF $F_2^b(x, Q^2)$,
$F_L(x, Q^2)$ and $F_L$ at fixed $W$ (see~\cite{JKLZ}). The UGD obtained from the fit to $F_2^c$
seems to overshoot the measured $F_2^b$ and $F_L$ at small $x$.  
New experimental data for $F_L (x, Q^2)$ but also more precise measurements of the heavy quark
structure functions are 
very important for a precise determination of the UGD.

\end{document}